\documentclass[prb,twocolumn]{revtex4-1}
\usepackage{graphicx}
\usepackage{dcolumn}
\usepackage{bm}
\usepackage{hyperref}
\usepackage{epstopdf}
\usepackage{epsfig}

\begin{document}

\title{Structure and magnetism of the $(2 \times 2)$-FeO(111) surface}
\author{I. Bernal-Villamil, S. Gallego} 
\email{sgallego@icmm.csic.es}
\affiliation{
Instituto de Ciencia de Materiales de Madrid, Consejo Superior de
Investigaciones Cient{\'{\i}}ficas, Cantoblanco, 28049 Madrid, Spain}

\date{\today}

\begin{abstract}
Based on {\it ab initio} calculations we determine the features and relative stability of different models
proposed to describe the $(2 \times 2)$-FeO(111) reconstruction. Our results suggest that both wurtzite and
spinel-like environments are possible, and their coexistence explains phenomena of biphase ordering.
The surface phase diagram reflects a competition of charge and magnetic compensation effects,
and reveals the important influence of the substrate on the final surface structure.
Though antiferromagnetic couplings are dominant, frustration and the delicate balance of surface and bulk
exchange interactions lead to a net surface magnetization that accounts for the large measured values.
\end{abstract}

\maketitle

\section{Introduction} 
The surfaces of binary Fe oxides are at the forefront of advanced technologies.
Because of their biocompatibility, they occupy a unique position in nanomagnetism,
with applications in areas as diverse as spintronics and biomedicine \cite{Lin-2015,Tartaj-2011}.
But they are also important due to their catalytic activity, in the design of gas sensor devices and for 
hydrogen storage \cite{Freund-2007,VietLong-2015,Otsuka-2003}, which confers increasing interest 
to the detailed knowledge of their structural and electronic properties under different environments.
Among these surfaces, those of FeO have an additional interest to design exchange bias nanocomposites \cite{Skumryev-2003,Sun-2012}.
Particularly, the FeO(111) termination is a good candidate to achieve large perpendicular magnetic anisotropy
\cite{Bechstedt-2015}
and has already been used as support of hexagonal two-dimensional materials like graphene \cite{Dahal-2015}.
%

Previous studies of the FeO(111) surface have led a considerable dispersion of results, 
partly due to the difficulties to prepare high quality samples.
On one hand, the bulk form only exists at ambient conditions with a significant 5-15$\%$ of Fe vacancies, and these vacancies 
tend to organize in different arrangements of local spinel-like clusters \cite{SilviaPRB2014}. 
Though Fe$_{1-x}$O samples can be stabilized by fast quenching, it is difficult to discern these 
clusters from Fe$_3$O$_4$ inclusions, which complicates their characterization.
On the other hand,  the phase diagram of binary Fe oxides can be strongly altered under reduced dimensions.
As a result, the preparation conditions in epitaxial growth
are crucial to achieve phase selectivity \cite{Gao-1998,Ketteler-2003,Monti-2012}.
In particular, while bulk FeO spontaneously decomposes into Fe$_3$O$_4$ and Fe, 
a precursor FeO layer forms during growth of Fe$_3$O$_4$(111) on most substrates, and often remains at the 
substrate/Fe$_3$O$_4$ interface \cite{Bernal-2015}. 
Also at the Fe$_3$O$_4$(111) surface a two-dimensional superlattice of Fe$_3$O$_4$
and FeO islands forms under low oxygen pressures ({\it biphase ordering}),
that under further reduction transforms to only FeO islands covering the Fe$_3$O$_4$ substrate 
 \cite{Condon-1997}. 

The tendency of stoichiometric FeO(111) to become stable in the ultrathin limit has been evidenced in numerous 
studies of mono- and bi-layer FeO(111) films on different supports.
Most works have been performed on Pt(111) \cite{Vurens-1992,Schedel-1995,Kim-1997}, but the use of other metals
and substrate orientations, including Pt(100) \cite{Vurens-1992}, Pd(111) \cite{Zeuthen-2013}, Pd(100) 
\cite{Kuhness-2016}, Ag(111) \cite{Waddill-2005}, Ag(100) \cite{Bruns-2014,Merte-2015}, Cu(110) \cite{Yagasaki-1994}
or Ru(0001) \cite{IreneJPc2013}, and even of different oxides such as YSZ(001) \cite{KellogSS2013} or 
$\alpha$-Al$_2$O$_3$(0001) \cite{Gota-2000} has been reported.
Though the substrate introduces subtle variations that manifest for example in the reactivity of the ultrathin
film, in all cases the Fe oxide seems to adopt an O-ended surface and a 
structure similar to the bulk, with a slightly expanded in-plane lattice and dominant
antiferromagnetic order \cite{Giordano-2007,Zhang-2009}.
Different attempts have been performed to overcome the bilayer thickness limit and grow thicker FeO films
\cite{Waddill-2005,SpiridisPRB2012,Yagasaki-1994,Corneille-1995}. 
The record thickness corresponds to 8 nm thick high-quality, stoichiometric FeO films with bulk-like properties 
obtained under reducing conditions on Ru(0001) \cite{LauraFeO}.
These films show a $(1 \times 1)$ O-termination that requires a wurtzite (WZ) stacking at the outermost 
surface layers, while preserving the AF-II magnetic order of FeO (namely, an alternance of ferromagnetic
Fe planes with opposite spin orientation along the $[111]$ axis).

But eventhough a $(1 \times 1)$ pattern has been found at slightly thinner FeO films on different
substrates \cite{Liu-2012,Merte-2015},
reconstructions emerge at higher O pressures: a $\sqrt{3} \times \sqrt{3} R 30^{\circ}$ surface linked 
to a transformation to $\alpha$-Fe$_2$O$_3$(0001) \cite{Galloway-1994}, and a poorly understood $(2 \times 2)$ structure 
\cite{Gao-1998,Cappus-1995,Mori-2005,Weiss-1993}. Different possibilities have been proposed to explain the
$(2 \times 2)$ symmetry, including stacking faults, an octopolar termination or the initial transformation to 
spinel Fe$_3$O$_4$(111), but its actual origin remains unclear.
The Fe valence is used to discriminate between FeO (with only Fe$^{2+}$) and Fe$_3$O$_4$ (with also Fe$^{3+}$),
but we have recently shown that Fe$^{3+}$ states may emerge even at the ideal rock-salt (RS) bulk truncation \cite{LauraFeO}.
As a further complication, sometimes the $(2 \times 2)$ symmetry corresponds to an actual transformation of
the entire film to Fe$_3$O$_4$ \cite{Kim-2000}. At present, there is no information about the electronic features 
of the spinel or octopolar geometries at the surface of a FeO(111) film.
More intriguing, for thin enough films grown on Fe(110), the $(2 \times 2)$ pattern can be accompanied by the 
emergence of ferromagnetism of high magnetization and ordering temperature \cite{Mori-2005}, 
temptatively explained assuming a model with significant 
Fe excess at the surface (see figure \ref{fmodels}b). But such model is contrary to the common evidence of 
an O-rich termination and to the fact that FeO is the end-phase in the Fe-rich limit.
Studies on nanoscaled films of FeO(111) have also measured a net magnetization that may survive at
high temperatures, but the magnetism seems to be linked to buried nanostructures or interface effects,
with no contribution from the bare surface \cite{Beach-2003,Couet-2009,SpiridisPRB2012}.

In the present work we address the detailed investigation of the $(2 \times 2)$-FeO(111) surface based
on {\it ab initio} calculations that continue our previous study of the unreconstructed termination \cite{LauraFeO}. 
Our purpose is to determine the accuracy of the models proposed from the experiments to describe
the $(2 \times 2)$ reconstruction, based both on their relative stability and their ability to reproduce
the measured properties.
Moreover, as we will show, our results account for the existence of biphase ordering phenomena.

\section{Methods} 
The conditions of the calculations have been explained elsewhere \cite{LauraFeO}. 
We use the density functional theory as implemented in the VASP code \cite{vasp}, based on the projector 
augmented wave (PAW) method to consider the core electrons. The exchange correlation part is described with the
Perdew-Burke-Ernzerhof parametrization of the generalized gradient approximation modified for solids (PBEsol),
including an effective local $U-J$ term following the Dudarev approach. The value $U-J=4$ eV has been chosen 
after calculations of bulk FeO \cite{SilviaPRB2014}.

We have modelled O-ended $(2 \times 2)$-FeO(111) slabs with the RS lattice of bulk FeO and a thickness of
9-10 planes, supported on a Ru(0001) layer and including a vacuum region of 14 $\AA$.
For all models and magnetic couplings, the atomic positions of the 5 outermost surface layers have been fully
relaxed using a $5 \times 5 \times 1$ Monhorst-Pack sampling of the Brillouin Zone until 
the forces on the atoms were below 0.01 eV$\AA^{-1}$. Convergence in the electronic properties and relative 
stabilities were then obtained with $9 \times 9 \times 2$ k-samplings. 

%
\begin{figure}[thbp]
\begin{center}
\includegraphics[width=\columnwidth,clip]{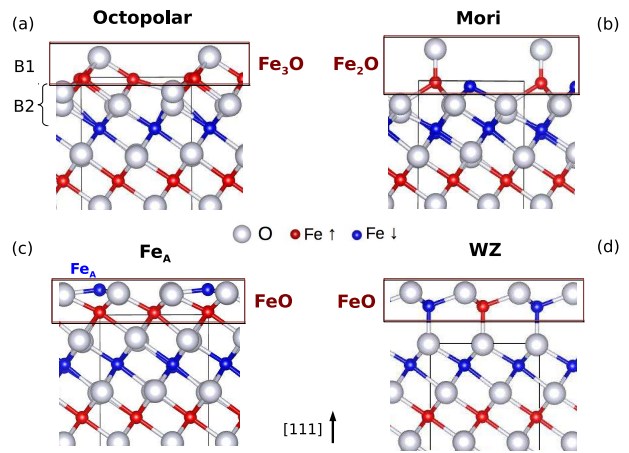}
\caption{(Color online).
Side view of the surface models considered, represented under their most stable structural and magnetic
configurations, and indicating the stoichiometry of the squared region B1 (see text for details). 
The grey lines in the back are a guide to delimit the size of the $(2 \times 2)$ unit cell.
\label{fmodels}}
\end{center}
\end{figure}
Figure \ref{fmodels} shows the different surface models considered by us. 
The surface region comprised by the four outermost planes includes a subsurface O-Fe bilayer (labelled B2) 
that resembles the bulk, 
and a surface O-Fe bilayer (B1, squared in the figure) that contains all relevant modifications.
Besides the models in the figure, we have also explored the possibility of a local hcp stacking at B1,
as proposed from analysis of low energy electron diffraction (LEED) beam profiles \cite{Weiss-1993}.
Such stacking increases the energy over the ideal RS termination over 1 eV. A WZ sequence would
also provide only two types of stacking sites at the surface region comprised by B1+B2, and can be at the origin
of the results in Ref. \cite{Weiss-1993}.

Both the octopolar\cite{Wolf-1992} and Mori\cite{Mori-2005} structures are based on partial removal of O atoms at B1, 
resulting in surface stoichiometries with an unlikely Fe excess. 
The Fe$_A$ model departs from the natural evolution of the rocksalt FeO(111)
stacking to a spinel Fe$_3$O$_4$(111) structure, shifting one Fe cation from the outermost Fe layer to
a tetrahedral coordination site (Fe$_A$) above the surface O plane. 
It also represents locally the surface environment corresponding to a 4:1 cluster of Fe vacancies \cite{SilviaPRB2014}.
As evidenced in the figure, the position of this Fe$_A$ cation relaxes to become almost coplanar to O, and 
the 1:1 Fe:O ratio is preserved at B1.
Finally, we have considered 
a WZ termination \cite{LauraFeO} where the $(2 \times 2)$ symmetry is introduced by the 
presence of in-plane antiferromagnetic couplings.
The electronic properties of the $(1 \times 1)$ WZ surface were already detailed in ref. \cite{LauraFeO}, 
and they are not significantly modified by the introduction of in-plane antiferromagnetism at B1.
Except for this WZ model, 
bulk electronic features are essentially recovered at B2 in all cases.

\section{Magnetic configurations} 
\begin{figure}[thbp]
\begin{center}
\includegraphics[width=0.8\columnwidth,clip]{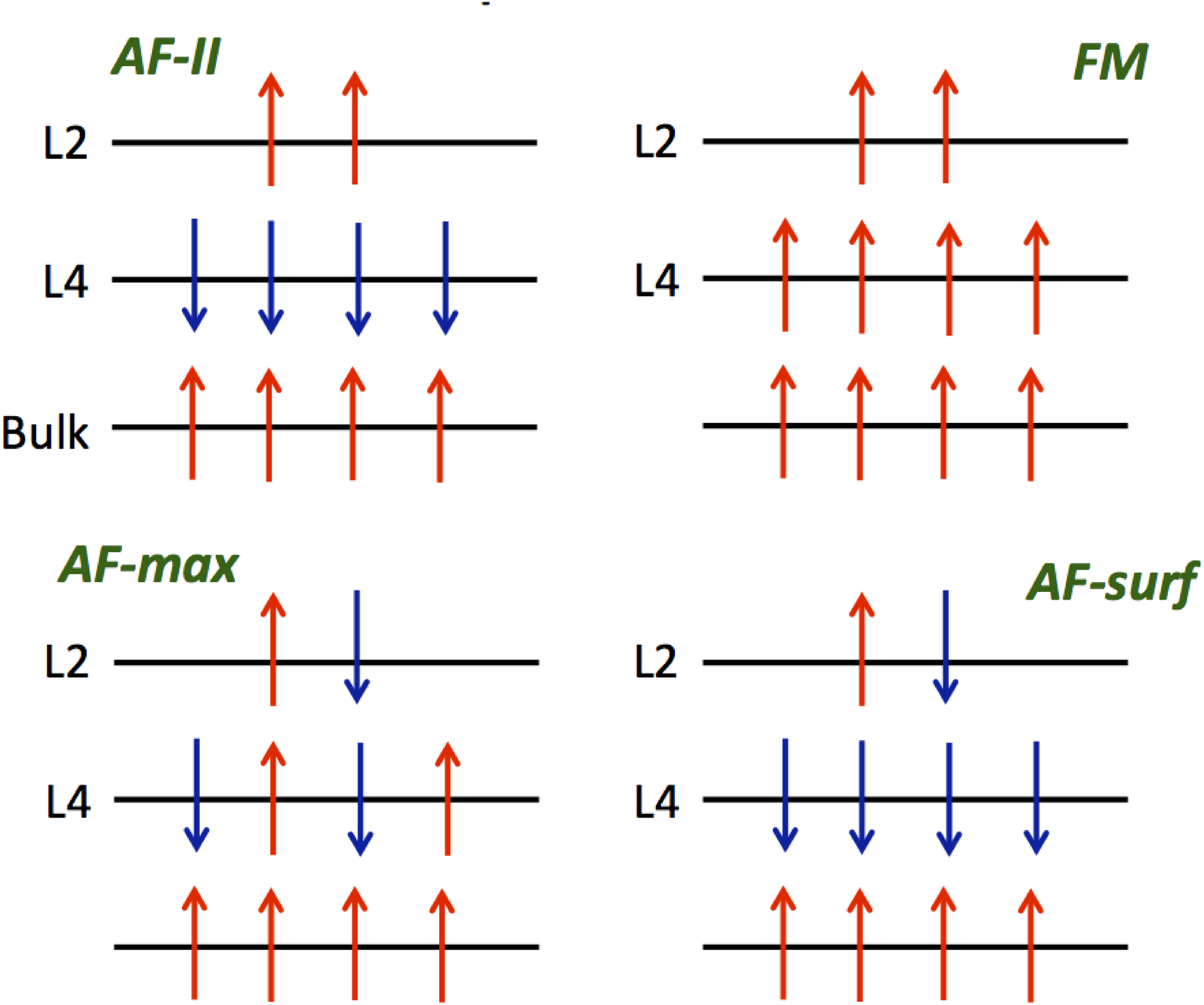}
\caption{(Color online).
Schematic representation of the different types of magnetic coupling considered for Fe atoms at B1 and B2.
\label{fmag}}
\end{center}
\end{figure}
For these structural models we have explored all possible collinear magnetic couplings between Fe atoms at B1 and B2 
compatible with the $(2 \times 2)$ symmetry, while keeping the AF-II order at the layers below.
Extending modifications of the AF-II order beyond B2 increased the total energy of the system.
A weak non-collinear magnetic component has been reported at the FeO(100) surface \cite{Bechstedt-2015}, 
but it emerges from the competition between the surface magnetic easy axis -that favors a 
perpendicular (to the surface) magnetization- and the bulk one -that lies along the (111) direction.
Such competition is not expected at the (111) surface, where a perpendicular magnetic component would 
align with the bulk easy axis. Furthermore, while evidence of non-collinear magnetic order has been
found at bulk CoO, the opposite holds for FeO \cite{Tomiyasu-2004}, justifying our approach.

The representative inequivalent configurations are shown in figure \ref{fmag}.
Though each configuration was allowed to relax independently, their structural differences
for a fixed surface model are minor, and do not alter the trends of relative stabilities.
It is important to keep in
mind that we are dealing with a magnetically frustrated system close to a charge instability, that 
under certain conditions may not even converge to a stable ground state (e.g. full ferromagnetism at B1 and B2).
Though under this situation we cannot provide a quantitative energy barrier, this
indicates that the explored solution is not favorable for the system.

The most stable magnetic solutions are those represented in figure \ref{fmodels}.
In the particular case of Mori model, we found as the stable magnetic order the same proposed from the 
experiments, the AF-surf configuration of figure \ref{fmag}, that lowers the energy with respect to the 
AF-II and AF-max couplings by 260 meV and 140 meV, respectively.
Analysis of all our results allows us to extract the following
conclusions: (i) shifting the AF-II order beyond B1 is always unfavorable; 
(ii) antiferromagnetic couplings dominate at the surface region, but the balance of in-plane and inter-layer 
exchange interactions at B1 and B2 depends on the detailed surface geometry and composition. 
The result is the continuity of the AF-II order under the octopolar
and Fe$_A$ terminations, while in-plane antiferromagnetism sets in at B1 for the rest of cases.
We would like to remark that we found such in-plane order to be favorable also under a RS surface termination, 
but the surface WZ stacking continues to be preferred over the RS one. 

\section{Results and discussion} 
\begin{figure}[thbp]
\begin{center}
\includegraphics[width=\columnwidth,clip]{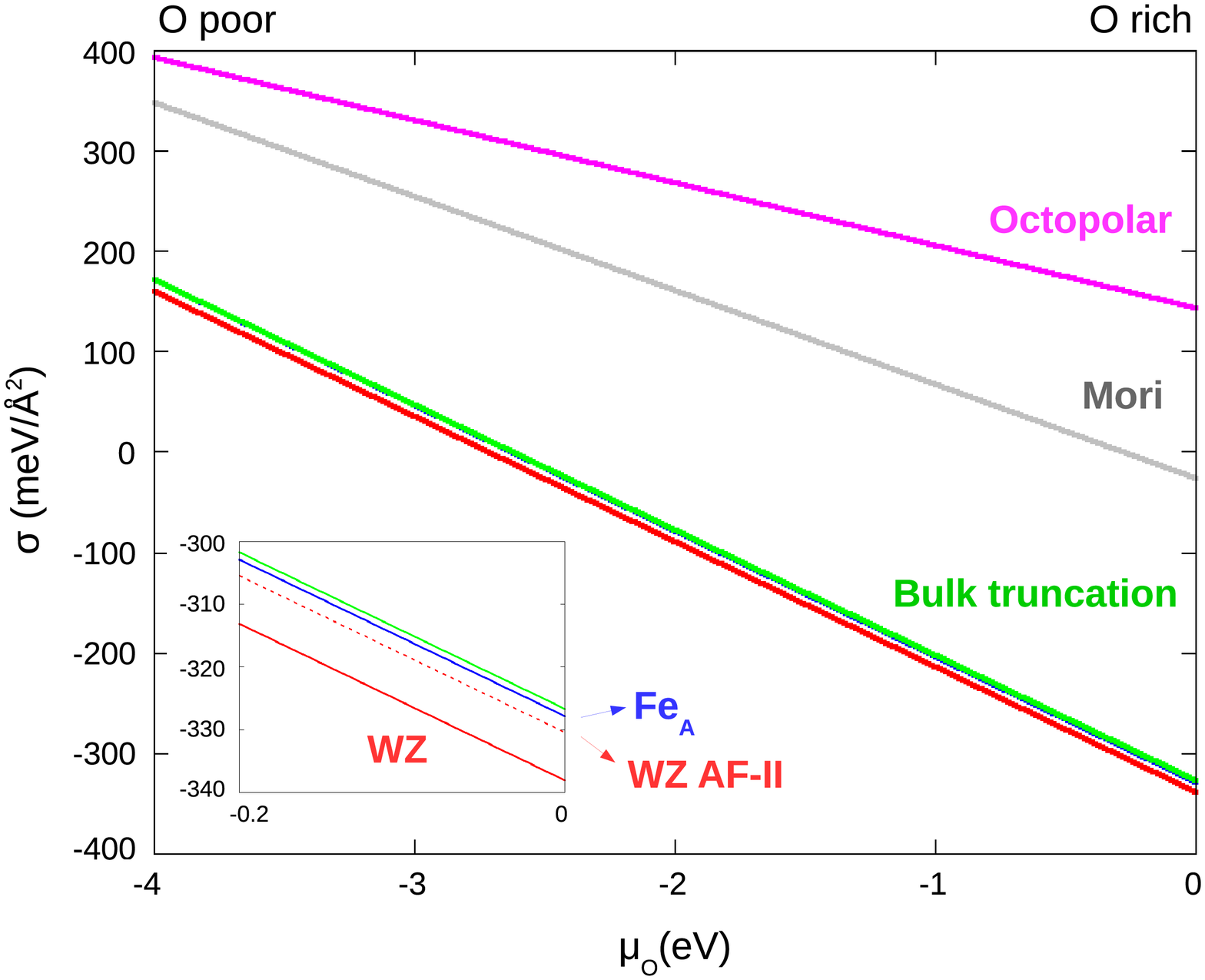}
\caption{(Color online).
Surface energy ($\sigma$) of the models in figure \protect\ref{fmodels} as a function of the O chemical 
potential ($\mu_O$). 
The RS (bulk truncation) and WZ terminations under AF-II order are also shown. 
The inset is a zoom of the lower right corner of the graph.
\label{fener}}
\end{center}
\end{figure}
Once we have the ground state magnetic configurations, we can determine the relative stability of the different surface models.
The identical stoichiometry of the Fe$_A$ and WZ structures allows direct comparison of their total
energies. However, consideration of the Mori and octopolar geometries
requires to take into account variations of the chemical potentials. In order to do so, the surface energy ($\sigma$) 
of all models has been computed following standard thermodynamic approximations \cite{Reuter-2001,LauraFeO}.
Since we have constructed asymmetric slabs supported on Ru and we are interested only in the O-ended surface, we refer 
the total energies to the common quantity $N_{Ru}\mu_{Ru}$, with $\mu_{Ru}$ extracted from the total energy of bulk Ru 
\cite{Franchini-2006}.
The results are shown in figure \ref{fener}, where the lowest surface energy corresponds to the most stable situation. 
As a reference, also the AF-II solutions for the RS bulk truncation and the WZ termination are shown.
It is evident that, as occured at the unreconstructed surface, the WZ stacking is always favored.
Also, that the Mori and octopolar models can be discarded even in the O poor limit.
On the other hand, the Fe$_A$ solution is close in energy to the bulk truncation and the WZ model, particularly 
when all are kept under a common AF-II magnetic order. 
These reduced energy differences explain the 
ease to transform the FeO(111) structure to a spinel-like surface, and the emergence of biphase ordering. 
%
%

%
\begin{table}[ht]
\begin{center}
\caption{
Mean Bader charge (Q) of the O and Fe atoms at B1, and net magnetization (in $\mu_B$ per unit cell) at
both B1 (M$_{B1}$) and the entire surface region (M$_{B1+B2}$) for the structures in figure \protect\ref{fmodels}. 
Corresponding values for the ideal AF-II RS bulk truncation are also shown.
In the Fe$_A$ model, the Q(Fe) value of the Fe$_A$ atom is distinguished from the rest.
\label{table}}
\renewcommand{\arraystretch}{1.5} 
\renewcommand{\tabcolsep}{0.4pc} 
\begin{tabular}{ccccc}
\hline \hline
Model & Q(O) & Q(Fe) & M$_{B1}$ & M$_{B1+B2}$ \\
\hline
 Octopolar        & 7.18 &       6.79    &  11.3  &  3.3  \\
 Mori             & 6.80 &       6.67    &   0.8  & 15.0  \\
 Fe$_A$           & 7.08 & 6.48$_A$/6.31 &   9.3  &  5.1  \\
 WZ               & 7.09 &       6.55    &   0.3  & 15.7  \\
 Bulk truncation  & 7.06 &       6.31    &  18.7  &  4.4  \\
\hline \hline
\end{tabular}
\end{center}
\end{table}
To better understand this surface phase diagram, we explore in detail the properties of the
FeO(111) reconstruction inherent to the different models.
An important aspect to take into account regarding the stabilization of the surface 
is the ability to compensate the loss of donor charge due to O bond breaking.
Table \ref{table} shows the mean Bader charges of the O and Fe atoms at B1 for all cases.
From the values for O, it is evident that Mori model leaves a high O charge deficiency, making it quite unplausible. 
The most efficient charge compensation is provided by the octopolar geometry, while
the rest of situations represent a slight improvement over the ideal bulk truncation.
But the large surface energy of the octopolar solution suggests that effects beyond mere charge 
compensation must influence the actual $(2 \times 2)$-FeO(111) structure.
Regarding the Fe charge, it contains information about the valence state of the surface Fe atoms 
\cite{LauraFeO,SilviaPRB2014}: values above 6.5 are linked to Fe$^{2+}$, typical of FeO,
while lower values represent a change to Fe$^{3+}$.
From the table, Fe$^{3+}$ only exists at the Fe$_A$ model, not only at the tetrahedral Fe$_A$ site, but at
the entire B1 bilayer. This serves to discard proposals of an octopolar geometry for FeO films grown on Pt(111),
where Fe$^{3+}$ states have been measured \cite{Galloway-1994}.
On the other hand, only Fe$^{2+}$ contributions have been identified at ultrathin films grown on Fe(110) \cite{Cappus-1995},
that would be compatible with all models except Fe$_A$.
Thus, analysis of the surface charge supports that the $(2 \times 2)$-FeO(111) surface admits multiple solutions,
in good agreement with the relative stabilities in figure \ref{fener}.
Furthermore, it reveals a substantial influence of the substrate on the surface phase diagram, 
opposite to the unreconstructed termination.

Further insights can be obtained from the surface magnetization arising from the different models.
Previously we showed that
antiferromagnetic interactions dominate at the outermost layers. However, this does not
necessarily imply the lack of a net surface magnetization.
%
This is best understood regarding the rightmost columns of table \ref{table}, where we have compiled 
the sum of individual atomic magnetic moments
at both B1 and the entire surface region (B1+B2) for the stable magnetic solutions in figure \ref{fmodels}.
Even without any reconstruction, creation of the surface 
causes an uncompensated magnetization \cite{LauraFeO}, reflected in the data for the bulk truncation.
Its value is enlarged by the
slight enhancement of the Fe magnetic moments at the surface layer, and also by the increased magnetization induced in
the outermost O atoms, that are only bonded to Fe atoms with one spin orientation.
These effects are also present under the $(2 \times 2)$ reconstruction, though depending on the surface structure, 
the individual Fe contributions at B1 may sum up or not. 
For cases that preserve the AF-II order, the magnetization at B1 is large, and even after adding
the contribution from B2 it remains significant, with the largest value for the Fe$_A$ case.
Oppositely, when in-plane antiferromagnetism exists, as occurs for the Mori and WZ models, 
the magnetization at B1 is almost negligible, while the sum of B1 and B2 is very large.
Interestingly, the maximum value does not correspond to Mori solution, but to the more stable WZ termination.
In summary, both the WZ and Fe$_A$ structures can explain the existence of a large magnetization at the surface that
surpasses that of a simple bulk truncation. Though our calculations do not address the estimation of the magnetic 
ordering temperature, the moderate magnetic energy differences suggest that the high measured values
arise from substrate induced effects in the ultrathin limit, as also pointed out from the experimental evidence \cite{Mori-2005}.

\section{Conclusions} 
We conclude that the $(2 \times 2)$ FeO(111) surface is a manifestation
of different surface structures, whose relative stability is largely conditioned by the choice of the substrate
and the preparation conditions. On one hand, it may be due to a local spinel structure, that either allocates Fe defect 
clusters or initiates a transition to Fe$_3$O$_4$(111). On the other, it may hold a WZ termination similarly to
the unreconstructed surface. Both structures are based on the introduction of local tetrahedral
environments at the outermost layers, but can be distinguished by their distinct signatures concerning
the presence of Fe$^{3+}$ states and the distribution of magnetic moments.
The coexistence of both solutions within a narrow energy window explains the origin of biphase ordering,
and makes room for the large influence of the substrate on the final structure.

The surface phase diagram obtained here reveals a subtle and complex role of magnetic interactions,
a situation similar to that found in bulk FeO \cite{SilviaPRB2014}.
While our results clearly discard a ferromagnetic solution, magnetic uncompensation
gives rise to large values of the surface magnetization.
%
A further complication that deserves to be considered is the influence of the distribution 
of Fe vacancies at the inner oxide layers, that could alter the surface magnetic properties.
However, for the ultrathin films where the $(2 \times 2)$-FeO(111) symmetry has been measured,
good stoichiometry at the inner layers can be assumed.
As a final consideration, in principle the surface magnetization reported here should
be distinguished from the magnetization measured at ultrathin FeO(111) films of 1-2 nm thickness 
\cite{SpiridisPRB2012},
more likely arising from the evolution of the in-plane antiferromagnetism of the monolayer
towards the layered bulk-like AF-II order. However, a non-zero surface contribution may exist, and
should be taken into account to explain the complex spectral magnetic features.

Our work represents a first step towards the understanding of the FeO(111) surface phase diagram.
It serves to rationalize the interpretation of the experiments, providing a reference
to identify intrinsic contributions arising from a defect free FeO film. This can be used as the starting
point of an ambitious molecular dynamics study directly incorporating substrate and temperature effects.
The complete picture still remains a challenging task, complicated by the ease to trigger the mutual transformation 
between the different binary Fe oxide phases at the nanoscale, largely conditioned by both external 
parameters (such as the choice of the substrate or the preparation conditions) and the presence of 
intrinsic defects or disorder.

\section*{Acknowledgments}
This research was supported by the Spanish Ministry of Economy and Competitiveness (MINECO) through grant 
MAT2012-38045-C04-04. I.B. acknowledges CSIC for the JAE program. The VESTA package\cite{VESTA} has been used 
to create some of the original figures of this article.

\end{document}